%% file: main.tex
\documentclass[11pt,letterpaper]{article}

\usepackage[utf8]{inputenc}
\usepackage[numbers,compress]{natbib}
\usepackage[english]{babel}

\usepackage{verbatim}
\usepackage{amsthm}
\usepackage{amsmath}
\usepackage[ruled,vlined]{algorithm2e}

\usepackage{booktabs}
\usepackage{enumerate}
\usepackage{graphics}
\usepackage{amssymb}
\usepackage{latexsym}
\usepackage{bbm}
\usepackage{needspace}
\usepackage{color, colortbl}
\usepackage[english]{babel}
\usepackage{tikz}
\usepackage{caption}
\usepackage{subcaption}
\usetikzlibrary{arrows,automata}
\usetikzlibrary{positioning}
\usepackage{filecontents}
\usepackage{mathtools}
\usepackage{multirow}
\usepackage{booktabs}

\usepackage[margin=1.0in]{geometry}
\theoremstyle{plain}

\newtheorem{theorem}{Theorem}

\newtheorem{definition}{Definition}
\newtheorem{problem}{Problem}
\newtheorem{assume}{Assumption}

\newtheorem{remark}{Remark}

\allowdisplaybreaks

\title{Learning Region of Attraction for Nonlinear Systems}

\author{Shaoru Chen, Mahyar Fazlyab, Manfred Morari, George J. Pappas, Victor M. Preciado
\thanks{Shaoru Chen, Manfred Morari, George J. Pappas, and Victor M. Preciado are with the Department of Electrical and Systems Engineering, University of Pennsylvania. Email: {srchen, morari, pappasg, preciado}@seas.upenn.edu. Mahyar Fazlyab is with the Mathematical Institute for Data Science, Johns Hopkins University. Email: mahyarfazlyab@jhu.edu.

This work was partially supported by the AFOSR grant FA9550-19-1-0265 (Assured Autonomy in Contested Environments).
}
}

\date{}
\begin{document}

\pagestyle{plain}
\maketitle

\begin{abstract}
\input{abstract}

\end{abstract}

\section{Introduction}
\input{introduction}

\section{Problem Statement}
\input{prob_statement}

\section{Parameterization of Robust Lyapunov functions}
\input{learning_Lyap}

\section{Sampling-based Lyapunov function synthesis}
\input{synthesis}

\section{Approximation of dynamical systems by neural networks}
\input{err_bound_assumption}

\section{Numerical examples}
\input{simulation}

\section{Conclusion}
\input{conclusion}

%\begin{appendices}
%	\section{}\label{appen:proof}
%	\input{appendixproof}
%\end{appendices}

\small
\bibliographystyle{IEEEtran}
\bibliography{reference}

\end{document}

%% file: abstract.tex
Estimating the region of attraction (ROA) of general nonlinear autonomous systems remains a challenging problem and requires a case-by-case analysis. Leveraging the universal approximation property of neural networks, in this paper, we propose a counterexample-guided method to estimate the ROA of general nonlinear dynamical systems provided that they can be approximated by piecewise linear neural networks and that the approximation error can be bounded. Specifically, our method searches for robust Lyapunov functions using counterexamples, i.e., the states at which the Lyapunov conditions fail. We generate the counterexamples using Mixed-Integer Quadratic Programming. Our method is guaranteed to find a robust Lyapunov function in the parameterized function class, if exists, after collecting a finite number of counterexamples. We illustrate our method through numerical examples.

%% file: introduction.tex
% introduction
%
Certifying the stability of nonlinear dynamical systems described by differential or difference equations is a long-standing fundamental problem in control theory. A general method for proving stability is the method of Lyapunov whereby one attempts to find, from a hypothesis class, a positive definite function that decreases along the trajectories of the dynamical system~\cite{haddad2011nonlinear, giesl2015review}. Once a Lyapunov function is found, its level sets can be used to estimate the region of attraction (ROA) of the equilibrium, which is a useful tool in constructing safety certificates~\cite{berkenkamp2016safe, wang2018permissive}.

In general, in order to find Lyapunov functions for nonlinear systems, one needs to solve a non-convex optimization problem. To avoid intractable non-convex optimization problems, there has been an increasing interest in developing data-driven and machine-learning inspired methods for stability analysis. For example, the authors in~\cite{richards2018lyapunov} parameterize the Lyapunov function as a neural network and train it through stochastic gradient descent to learn the largest ROA of a nonlinear system. Giesl et al.~\cite{giesl2016approximation} learn a Lyapunov function as a reproducing kernel Hilbert space predictor from noisy trajectories of a nonlinear system. Boffi et al.~\cite{boffi2020learning} treat Lyapunov function synthesis as a machine learning problem and provide generalization error bounds for the learned Lyapunov function. 

Although success examples have been demonstrated for the learning-based methods, the guarantees on the stability certificates are often of probabilistic nature since only sampled trajectory data are used. On the other hand, feedforward neural network or recurrent neural network models obtained from learning-based system identification are shown to have a strong predictive power~\cite{ogunmolu2016nonlinear} and can approximate complex nonlinear dynamics with small errors. This offers an opportunity to obtain deterministic stability guarantees on nonlinear systems by analyzing their neural network approximations with the model errors taken into account. However, how to analyze the stability of uncertain neural network dynamics remains a challenge with only limited work available~\cite{chen2021learning, yin2020stability}. 

In this paper, we propose a sampling-based method for constructing robust Lyapunov functions (see Theorem~\ref{thm:Lyap} for definition) for a class of uncertain dynamical systems with feedforward ReLU networks modeling the nominal dynamics over a compact region of interest (ROI). Our method consists of a learner, which is responsible for generating Lyapunov function candidates, and a verifier, which validates or rejects the proposed candidates with counterexamples to guide the search. More importantly, if a Lyapunov function exists in the parameterized Lyapunov function class, our method is guaranteed to find one in a finite number of steps. These properties are particularly desirable for analysis of uncertain neural network dynamical systems. For comparison, the convex relaxation based methods~\cite{yin2020stability} easily become excessively conservative as the size of the neural network or the model uncertainty increases. The counterexample-guided Lyapunov function synthesis methods in~\cite{ahmed2020automated, abate2020formal, chang2020neural} do not have finite-step termination guarantees which means that the number of calls to the computationally expensive verifier cannot be bounded. Our method is inspired by~\cite{chen2021learning} which proposes a sampling-based method for stability analysis of autonomous hybrid systems. We extend the method in~\cite{chen2021learning} to handle uncertain dynamical systems and demonstrate the potential of our proposed method to automate Lyapunov function synthesis for nonlinear systems through their neural network dynamics approximations.

We state the assumption on the uncertain neural network dynamical system and the problem formulation in Section~\ref{sec:formulation}. In Section~\ref{sec:Lyapunov_synthesis}, the parameterization of robust Lyapunov function is introduced and a sampling-based method to synthesize a robust Lyapunov function is presented in Section~\ref{sec:synthesis}. The practical use of the proposed framework is discussed in Section~\ref{sec:applicability} with numerical examples given in Section~\ref{sec:simulation} to demonstrate our method. Section~\ref{sec:conclusion} concludes the paper.

%By introducing a verifier which is capable of validating or rejecting a Lyapunov function candidate, the data-driven synthesis methods~\cite{ravanbakhsh2019learning, chang2020neural, ahmed2020automated} are able to provide provable guarantees on the synthesized stability certificate.

% In order to find Lyapunov functions for nonlinear systems, one needs to solve a non-convex optimization problem, the complexity of which depends on the system dynamics, the hypothesis class, and the ambient dimension. Many computational methods have been developed to construct Lyapunov functions for various classes of dynamical systems such as series expansion [REF], linear programming [REF], linear matrix inequalities [REF], collocation methods [REF], algebraic methods [REF], set-theoretic methods [REF], and many others. To avoid intractable non-convex optimization problems for general dyanamical systms, there has been an increasing interest in developing data-driven and machine-learning inspired methods for stability analysis of dynamical systems [REFS].

%% file: prob_statement.tex
%problem statement
\label{sec:formulation}
Consider a discrete-time nonlinear dynamical system 
\begin{equation} \label{eq:sys_dyn}
x_+ = f(x)
\end{equation}
where $x \in \mathbb{R}^{n_x}$ is the state, $x_+$ denotes the state at the next time instance, and $f: \mathbb{R}^{n_x} \mapsto \mathbb{R}^{n_x}$ is a locally Lipschitz continuous function. We assume that system~\eqref{eq:sys_dyn} has an equilibrium $x = 0$, i.e., $0 = f(0)$, and is asymptotically stable at the origin. Verifying that system~\eqref{eq:sys_dyn} is locally stable around the origin can be done by analyzing the linearization of the system at $x=0$ if it exists. Of more practical use is estimating the region of attraction of the origin.

\begin{definition}[Region of attraction]
The region of attraction for system~\eqref{eq:sys_dyn} at the origin is $\mathcal{O} = \{x_0 \vert \lim_{k \rightarrow \infty} x_k = 0 \}$ where $x_k$ denotes the state of the system at time $k=0,1,2,\cdots$. 
\end{definition}

The ROA characterizes how much the system can be perturbed from $x = 0$ without diverging. Therefore, the ROA is often used to describe the safety of nonlinear systems. In this work, we are also interested in the region where the convergence of system trajectories to a small neighborhood of the origin can be verified since this relaxed notion of ROA also suffices to show safety of nonlinear systems in practical applications.

Computing the exact ROA $\mathcal{O}$ for general nonlinear systems is infeasible and hence, one seeks to find an estimate $\tilde{\mathcal{O}}$ of ROA for specific classes of dynamical systems. In this work, we assume that the map $f$ can be approximated by a piecewise linear multi-layer neural network $f_{NN}$ and that the approximation error $f-f_{NN}$ can be bounded. Specifically, consider a neural network with ReLU\footnotemark activation:
\footnotetext{ReLU (rectified linear unit) activation function is defined by $\phi(x) = \max(0, x)$ for $x\in\mathbb{R}^n$ where the maximum is applied entry-wise.}
\begin{equation} \label{eq:network}
	\begin{aligned}
		z_0 &= x\\
		z_{\ell+1} &= \max(W_\ell z_\ell + b_\ell, 0), \quad \ell = 0, \cdots, L -1 \\
		f_{NN}(x) &= W_{L} z_{L} + b_L
	\end{aligned}
\end{equation}
where $z_0 = x \in \mathbb{R}^{n_0} \ (n_0 = n_x)$ is the input to the neural network, $z_{\ell+1} \in \mathbb{R}^{n_{\ell+1}}$ is the the output vector of the $(\ell+1)$-th hidden layer with $n_{\ell+1}$ neurons, $f_{NN}(x) \in \mathbb{R}^{n_{L+1}} \ (n_{L+1} = n_x)$ is the output of the neural network, and $W_\ell \in \mathbb{R}^{n_{\ell + 1} \times n_\ell}, b_\ell \in \mathbb{R}^{n_{\ell + 1}}$ are the weight matrix and the bias vector of the $(\ell+1)$-th hidden layer. 
\begin{assume} \label{assumption:nn approximation}
	For a given Schur stable matrix $A \in \mathbb{R}^{n_x \times n_x}$ and a compact region of interest (ROI) $\mathcal{X} \subset \mathbb{R}^{n_x}$ with $0 \in \text{int}(\mathcal{X})$, where $\text{int}(\mathcal{X})$ denotes the interior of $\mathcal{X}$, the map $f(x)$ can be decomposed as
	\begin{equation}\label{eq:nn_approx_dyn}
		f(x) = Ax + f_{NN}(x) + w(x), \quad \forall x \in \mathcal{X}
	\end{equation}
   where the approximation error $w(x) \in \mathbb{R}^{n_x}$ satisfies
   \begin{equation}\label{eq:approx_err_bd}
   	 \lVert w(x) \rVert_{\infty} \leq \gamma \lVert x \rVert_{\infty} + \delta, \quad \forall x \in \mathcal{X},
   \end{equation}
   for some known positive constants $\gamma,\delta$.
\end{assume}
Now consider the uncertain dynamical system
\begin{equation}\label{eq:uncertain_dyn}
	x_+ = \underbrace{Ax + f_{NN}(x)}_{:=\hat f(x)}+ w, \quad w \in \mathcal{W}(x)
\end{equation}
where $\mathcal{W}(x) = \{w \in \mathbb{R}^{n_x} \mid  \lVert w \rVert_{\infty} \leq \gamma \lVert x \rVert_{\infty} + \delta \}$ is a state dependent set that contains the ``disturbance'' $w$. Under Assumption~\ref{assumption:nn approximation}, the nonlinear dynamics $f(x)$ is contained in the uncertain dynamics class described by~\eqref{eq:uncertain_dyn} which enjoys a specific structure that can be exploited for stability analysis. The main problem considered in this paper is stated as follows.
\begin{definition}[Robust ROA]
Let $x_k$, $w_k$ denote the state and disturbance of the uncertain system~\eqref{eq:uncertain_dyn} at time $k=0, 1, \cdots$, respectively, and $B \in \mathcal{X}$ be a small neighborhood of the origin with $0 \in \text{int}(B)$. The robust ROA of the uncertain system~\eqref{eq:uncertain_dyn} is defined as $\mathcal{O} = \{ x_0 \vert \lim_{k \rightarrow \infty} x_k \in B, \forall w_k \in \mathcal{W}(x_k) \}$.
\end{definition}

\begin{problem}
Find an estimate of the robust ROA of the uncertain neural network dynamical system~\eqref{eq:uncertain_dyn}.
\end{problem}

In the Section~\ref{sec:Lyapunov_synthesis} and~\ref{sec:synthesis}, we propose a sampling-based method to synthesize a robust Lyapunov function which identifies an estimate of the robust ROA $\tilde{\mathcal{O}}$ for the uncertain system~\eqref{eq:uncertain_dyn}. By virtue of Assumption~\ref{assumption:nn approximation}, $\tilde{\mathcal{O}}$ would be a valid inner estimate of ROA for system~\eqref{eq:sys_dyn}. We postpone the discussion on the applicability of Assumption~\ref{assumption:nn approximation} to Section~\ref{sec:applicability}.

%% file: learning_Lyap.tex
% Lyapunov function synthesis
\label{sec:Lyapunov_synthesis}
%In~\cite{chen2021learning}, a sampling-based method is proposed to synthesize Lyapunov functions for autonomous hybrid systems. Inspired by~\cite{chen2021learning}, we propose an extension of this method to uncertain dynamical systems, and synthesize \emph{robust Lyapunov functions} that monotonically decrease for all possible realization of the uncertain system. Specifically, we propose an iterative algorithm that can exhaustively search over the space of robust Lyapunov function candidates and is guaranteed to find a robust Lyapunov function in a finite number of iterations if the set of robust Lyapunov functions is full-dimensional. Once a robust Lyapunov function is found, its sublevel sets encode an inner estimate of the ROA for the uncertain system~\eqref{eq:uncertain_dyn} and hence the original system~\eqref{eq:sys_dyn}. We explain the details of our algorithm in the next few subsections.

For the uncertain system~\eqref{eq:uncertain_dyn}, we aim to find a robust Lyapunov function that monotonically decreases for all possible realization of uncertainty. Once a robust Lyapunov function is found, its sublevel sets encode an inner estimate of the robust ROA for the uncertain system~\eqref{eq:uncertain_dyn}, and hence the original system~\eqref{eq:sys_dyn}. 

\subsection{Robust Lyapunov function}
For the uncertain system~\eqref{eq:uncertain_dyn} with the uncertainty set $\mathcal{W}(x)$, we aim to show that the system trajectories converge to a neighborhood $\mathcal{B}$ of the origin. The neighborhood $\mathcal{B}$ is chosen according to the practical need of verification. Let $B_r = \{ x\in \mathbb{R}^{n_x} \vert \lVert x \rVert_\infty \leq r \} \subset \mathcal{X}$ denote an $\ell_\infty$ norm ball with radius $r > 0$ contained inside the ROI. The robust Lyapunov function is defined in the following theorem. 

\begin{definition}[Successor set]
For the uncertain dynamics~\eqref{eq:uncertain_dyn}, we denote the successor set from a set $\mathcal{B}$ as $\text{succ}(\mathcal{B}) = \{y \in \mathbb{R}^{n_x} \vert \exists x \in \mathcal{B}, w \in \mathcal{W} \text{ s.t. } y = \hat{f}(x) + w\}$. 
\end{definition}
\begin{theorem} \label{thm:Lyap}
	Consider the discrete-time uncertain system~\eqref{eq:uncertain_dyn} and assume there exists a continuous function $V(x): \mathcal{X} \mapsto \mathbb{R}$ satisfying
	\begin{subequations}
		\label{eq:robust_Lyap_cond}
		\begin{align}
		\begin{split} \label{eq:robust_Lyap_cond_1}
		&V(x) > 0, \forall x \in \mathcal{X} \setminus B_r,
		\end{split} \\
		\begin{split} \label{eq:robust_Lyap_cond_2}
		& V(\hat{f}(x) + w) - V(x) < 0, \forall x \in \mathcal{X}\setminus B_r, \forall w \in \mathcal{W}(x).
		\end{split}
		\end{align}
	\end{subequations}
	Denote $\underline{\Omega}_{\text{succ}(B_r)}(V)$ the smallest sublevel set of $V(x)$ that contains the successor set of $B_r$ and $\bar{\Omega}_{\mathcal{X}}(V)$ the largest sublevel set of $V(x)$ that is contained in $\mathcal{X}$. Then for all initial states $x_0 \in \bar{\Omega}_{\mathcal{X}}(V)$ of the uncertain system~\eqref{eq:uncertain_dyn}, we have $\lim_{k \rightarrow \infty} x_k \in \underline{\Omega}_{\text{succ}(B_r)}(V)$.
\end{theorem}
\begin{proof}
All trajectories starting from $x_0 \in \bar{\Omega}_\mathcal{X}(V)$ reach $B_r$ in a finite number of steps since $V(x)$ is upper bounded in $\mathcal{X}$ and its decrease in each step is lower bounded by a positive number as long as $x \notin B_r$. Once the state $x_k$ reaches $B_r$, it is mapped to $x_{k+1} \in \text{succ}(B_r) \subset \underline{\Omega}_{\text{succ}(B_r)}(V)$ in the next step. Whether $x_{k+1} \notin B_r$ or $x_{k+1} \in B_r$, the trajectory starting from $x_{k+1}$ will always stay in $\underline{\Omega}_{\text{succ}(B_r)}(V)$.
\end{proof}
We call any $V(x)$ satisfying constraints~\eqref{eq:robust_Lyap_cond_1} and~\eqref{eq:robust_Lyap_cond_2} a \emph{robust Lyapunov function} and those satisfying constraint~\eqref{eq:robust_Lyap_cond_1} a robust Lyapunov function candidate. Theorem~\ref{thm:Lyap} states that a robust Lyapunov function $V(x)$ certifies the convergence of system trajectories under uncertainty to a neighborhood of the origin whose size and shape is jointly decided by $V(x)$ and $B_r$. The largest sublevel set of $V(x)$ contained in the ROI $\mathcal{X}$ gives an estimate of robust ROA for the uncertain system, and $\tilde{\mathcal{O}} = \bar{\Omega}_{\mathcal{X}}(V)$ is also an inner-approximation of ROA of the original nonlinear system~\eqref{eq:sys_dyn}. In the rest of the paper, we refer to robust Lyapunov functions simply as Lyapunov functions.

%With a small $B_r$, the verified attractive neighborhood tends to be also small, but synthesizing a robust Lyapunov function becomes more challenging. 

%In this work, we let parameterize the ROI $\mathcal{X}$ as a polytopic set
%\begin{equation}
%\mathcal{X} = \{x \in \mathbb{R}^{n_x} \vert F_\mathcal{X} x \leq h_\mathcal{X}\}
%\end{equation}
%and $B_r$ as a $\ell_\infty$ norm ball with radius $r > 0$. Obviously, we should let $r > \delta$ in order to find a robust Lyapunov function. Both $\mathcal{X}$ and $B_r$ are tuning parameters in our search for robust Lyapunov functions $V(x)$. In our experiments, we first simulate the system trajectories to `guess' the shape of the ROA and then use $\mathcal{X}$ and $B_r$ to encode such information. 

%Once a robust Lyapunov function $V(x)$ is found, its sublevel set inside $\mathcal{X}$ comprises an estimate of ROA for the uncertain system, i.e., we can choose $\tilde{\mathcal{O}} = \{x \vert V(x) \leq \alpha\}$ with some scalar $\alpha > 0$ such that $B_r \subset \tilde{\mathcal{O}} \subset \mathcal{X}$. Then for any $x_0 \in \tilde{\mathcal{O}}$, we have $\lim_{k \rightarrow \infty} x_k \in B_r$ for the uncertain system~\eqref{eq:uncertain_dyn} and hence the original system~\eqref{eq:sys_dyn}

\subsection{Lyapunov function parameterization}
To make the search for a Lyapunov function $V(x)$ tractable, we parameterize the Lyapunov function candidate as a quadratic function using predicted nominal trajectories of the uncertain system~\eqref{eq:uncertain_dyn} for a finite horizon as the basis. For a horizon of $k \geq 0$, the basis is given by 
\begin{equation*}
z_k(x) = [ x^\top \ \hat{f}^{(1)}(x)^\top \ {\hat{f}^{(2)}(x)}^\top \ \cdots \ {\hat{f}^{(k)}(x)}^\top ]^\top
\end{equation*}
where $\hat{f}^{(i+1)}(x) = \hat{f}(\hat{f}^{(i)}(x))$ for $i \geq 0$ and $\hat{f}^{(0)}(x) = x$. By denoting $\mathbb{S}^n$ the set of $n \times n$-dimensional symmetric matrices, and $\mathbb{S}^n_{+}$ ($\mathbb{S}^n_{++}$) the set of $n \times n$-dimensional positive semidefinite (definite) matrices, the Lyapunov function candidate of \emph{order $k$} is defined as
\begin{equation} \label{eq:V_order}
	V_k(x; P) = z_k^\top(x) P z_k(x), \quad P \in \mathbb{S}_{++}^{n_x}.
\end{equation}
This parameterization is inspired by the non-monotonic Lyapunov function~\cite{ahmadi2008non} and finite-step Lyapunov function~\cite{bobiti2016sampling} methods which construct Lyapunov function candidates using system states several steps ahead. By construction, $V_k(x;P)$ is positive definite and readily satisfies constraint~\eqref{eq:robust_Lyap_cond_1}. In addition, the parameterization of $V_k(x;P)$ has two desirable properties: (i) $V_k(x;P)$ is a linear function in $P$ when the state $x$ is fixed, and (ii) for $k \geq 1$, $V_k(x;P)$ is a piecewise quadratic function in $x$ and its complexity can be easily tuned by the order $k$. 

The first property indicates that the set of valid Lyapunov function candidates is convex. To show this, define the Lyapunov difference function as 
\begin{equation}
\Delta V_k(x, w, P) = V_k(\hat{f}(x) + w; P) - V_k(x; P).
\end{equation}
From the first property of $V_k(x;P)$, we know that $\Delta V_k(x, w, P)$ is also a linear function in $P$ when the state $x$ and uncertainty $w$ are fixed. We describe the set of valid Lyapunov functions in the parameter space as
\begin{equation} \label{eq:target_set}
\begin{aligned}
\mathcal{F} = \{ P \in \mathbb{S}^{n_x} \vert 0 \prec P \preceq I ,  \Delta V_k(x, w, P) < 0, \forall x \in \mathcal{X}\setminus B_r, \forall w \in \mathcal{W}(x)\}
\end{aligned}
\end{equation}
and denote $\mathcal{F}$ the \emph{target set}. It follows that $\mathcal{F}$ is a convex set since it is defined by two linear matrix inequalities and infinitely many linear inequalities in $P$. The convexity of $\mathcal{F}$ is essential for the design of our main algorithm to search for Lyapunov functions.

The second property follows from the fact that the ReLU network $\hat{f}(x)$ is a piecewise affine function and so are $\hat{f}^{(i)}(x)$. By tuning the order $k$, we can parameterize functions with high complexity, i.e., piecewise quadratic functions with a large amount of partitions, by using a relatively small number of parameters. 

%This is in spirit with the construction of non-monotonic Lyapunov functions~\cite{ahmadi2008non} which increase the complexity of Lyapunov function candidates by incorporating predicted system states several steps ahead. 

After identifying the set of Lyapunov functions $\mathcal{F}$, our goal becomes finding a feasible point in $\mathcal{F}$ or proves that $\mathcal{F}$ is empty with the given parameters. Although the set $\mathcal{F}$ is convex, finding a feasible point in it is challenging due to the infinitely many constraints in~\eqref{eq:target_set} and the complexity of the underlying uncertain neural network dynamics~\eqref{eq:uncertain_dyn}.

%% file: synthesis.tex
% Lyapunov function synthesis
\label{sec:synthesis}
In this section, we extend the sampling-based method in~\cite{chen2021learning} to synthesize a Lyapunov function for the uncertain system~\eqref{eq:uncertain_dyn} with neural network nominal dynamics. Specifically, the proposed method is guaranteed to find a Lyapunov function in the target set $\mathcal{F}$ in a finite number of iterations when $\mathcal{F}$ is full-dimensional, and is capable of providing an infeasibility certificate of $\mathcal{F}$ when $\mathcal{F}$ is empty. Similar to~\cite{chen2021learning}, we obtain this guarantee by sampling according to the analytic center cutting-plane method~\cite{nesterov1995cutting, atkinson1995cutting} from convex optimization. A step-by-step explanation is given below.

\subsection{Localization of the target set}
To handle the infinitely many linear constraints in~\eqref{eq:target_set}, we first observe that an over-approximation $\tilde{\mathcal{F}}$ of the target set $\mathcal{F}$ can be obtained by relaxing the Lyapunov difference constraint in~\eqref{eq:target_set} to hold only on a sample set with a finite number of state-disturbance pairs $\mathcal{S} = \{(x^S_i, w^S_i)\}_{i=1}^N$, i.e., 
\begin{equation*}
\begin{aligned}
\tilde{\mathcal{F}} = \{ P \in \mathbb{S}^{n_x} \vert 0 \prec P \preceq I , \Delta V_k(x, w, P) < 0,  \forall (x, w) \in \mathcal{S} \}
\end{aligned}
\end{equation*}
with samples $x^S_i \in \mathcal{X} \setminus B_r, w_i^S \in \mathcal{W}(x_i^S)$ for $i = 1,\cdots, N$. We call $\tilde{\mathcal{F}}$ a localization set of $\mathcal{F}$ and $\mathcal{F}\subseteq \tilde{\mathcal{F}}$ holds by definition. Since the localization set $\tilde{\mathcal{F}}$ is given by a finite number of convex inequalities, finding a feasible point in $\tilde{\mathcal{F}}$ or showing infeasibility of $\tilde{\mathcal{F}}$ can be done by solving a convex program efficiently. If for a sample set $\mathcal{S}$, the localization set $\tilde{\mathcal{F}}$ is empty, then we have a certificate that the target set $\mathcal{F}$ is empty. When the target set is nonempty, we can find a feasible point in $\mathcal{F}$ by iteratively expanding the sample set $\mathcal{S}$ and refining the localization set $\tilde{\mathcal{F}}$. However, expanding the sample set in an arbitrary or random way is not efficient or may not improve the over-approximation of $\mathcal{F}$; in order to obtain performance guarantees, we need to add samples that are informative at each iteration. This is done by alternating between a learner and a verifier to select the samples.

\subsection{Construction of the learner}
Given a sample set $\mathcal{S}$ and the corresponding localization set $\tilde{\mathcal{F}}$, the learner proposes a Lyapunov function candidate $V_k(x; P_{ac})$ with $P_{ac}$ as the analytic center of $\tilde{\mathcal{F}}$.

\begin{definition}[Analytic center~\cite{boyd2004convex}] \label{def:analytic_center}
	The analytic center $x_{ac}$ of a set of convex inequalities and linear equalities
	%\begin{equation*}
	$h_i(x) \leq 0, i = 1,\! \cdots\!, m, \ Fx = g$,
	%\end{equation*}	
	is defined as the solution of the convex problem
	\begin{equation*}
	\begin{aligned}
	\underset{x}{\mathrm{minimize}} & \quad - \sum_{i=1}^m  \log(-h_i(x))  \quad \text{subject to} \quad Fx = g.
	\end{aligned}
	\end{equation*}
\end{definition}

By the definition of analytic center, $P_{ac}$ is given by
\begin{equation}\label{eq:ac_optimization}
\begin{aligned}
P_{ac} := \underset{P}{\text{argmin}} \ -\sum_{(x, w)\in \mathcal{S}} \log( -\Delta V_k(x, w, P) ) - \log\det (I - P) - \log \det (P).
\end{aligned}
\end{equation}
where $- \log\det (I - P) - \log \det (P)$ is the barrier function of the set $\{P \vert 0 \prec P \prec I \}$. If problem~\eqref{eq:ac_optimization} is infeasible, the localization set $\tilde{\mathcal{F}}$ is empty and so is the target set $\mathcal{F}$. If problem~\eqref{eq:ac_optimization} is feasible, $V_k(x; P_{ac})$ is proposed by the learner as the Lyapunov function candidate based on the localization set $\tilde{\mathcal{F}}$ and is passed to the verifier.

\subsection{Construction of the verifier}
For the given Lyapunov function candidate $V_k(x;P_{ac})$ proposed by the learner, the verifier either validates $V_k(x;P_{ac})$ or rejects it by giving a counterexample $(x^*, w^*)$ at which $V_k(x;P_{ac})$ violates the Lyapunov difference condition~\eqref{eq:robust_Lyap_cond_2}, i.e., $\Delta V_k(x^*, w^*, P_{ac}) \geq 0$. This is done by solving the global optimization problem:
\begin{alignat}{2} \label{eq:verifier_formulation}
&p^* = \underset{x \in \mathcal{X} \setminus B_r, w \in \mathcal{W}(x)}{\mathrm{maximize}} \quad && \Delta V_k(x,w,P_{ac})
\end{alignat}
which is nonconvex. Denote $p^*$ the optimal value of~\eqref{eq:verifier_formulation} and $(x^*, w^*)$ the optimal solution. The verifier validates $V_k(x;P_{ac})$ if $p^* < 0$ and rejects $V_k(x;P_{ac})$ if $p^* \geq 0$ with $(x^*, w^*)$ as the counterexample. In the former case, we already have a valid Lyapunov function $V_k(x;P_{ac})$ for the uncertain system~\eqref{eq:uncertain_dyn}; in the latter case, the counterexample $(x^*, w^*)$ is added to the sample set $\mathcal{S}$ and the localization set $\tilde{\mathcal{F}}$ is updated. Then the learner generates a new Lyapunov function candidate based on the updated localization set. The alternation between the learner and the verifier is repeated until either a Lyapunov function in $\mathcal{F}$ is found or $\mathcal{F}$ is certified to be empty. We summarize our method in Algorithm~\ref{alg:ACCPM}.
\begin{algorithm}
	\KwData{Initial sample set $\mathcal{S}$}
	\KwResult{Infeasibility certificate or Lyapunov function}
	\While{True}{
		Update localization set $\tilde{\mathcal{F}}$ \\
		Call the learner \\
		\eIf{\eqref{eq:ac_optimization} is feasible}{Propose candidate $V_k(x;P_{ac})$ }{Return infeasibility certificate \\}
		Call the verifier \\
		\eIf{max~\eqref{eq:verifier_formulation} $< 0$}{Return Lyapunov function }{Generate counterexample $(x^*, w^*)$ \\
			Update sample set $\mathcal{S} = \mathcal{S} \cup \{(x^*, w^*)\}$}
	}
	\caption{Learning robust Lyapunov function}
	\label{alg:ACCPM}
\end{algorithm}
Solving a nonconvex optimization problem to global optimality is in general intractable, but for the uncertain neural network dynamical system~\eqref{eq:uncertain_dyn}, we are able to formulate problem~\eqref{eq:verifier_formulation} as a nonconvex mixed-integer quadratic program (MIQP) for which global optimization algorithms exists~\cite{belotti2013mixed, belotti2009branching}. In practice, Algorithm~\ref{alg:ACCPM} is easy to implement since solving nonconvex MIQPs to global optimality can be done in off-the-shell solvers such as Gurobi v9.0~\cite{gurobi}. Interested readers are referred to~\cite{chen2021learning} for more discussions on the complexity and solvability of MIQPs. Next, we show how to formulate problem~\eqref{eq:verifier_formulation} as an MIQP by exploiting the mixed-integer linear formulation of ReLU networks. 

The formulation of the MIQP is based on the fact that the feedforward ReLU network $f_{NN}(x)$ can be described by a set of mixed-integer linear (MIL) constraints. For the $(\ell+1)$-th activation layer in the ReLU network described in \eqref{eq:network}, let $\underline{m}^{\ell}$ and $\bar{m}^{\ell}$ be the element-wise lower and upper bounds on the input, i.e., $\underline{m}_{\ell} \leq W_{\ell} z_{\ell} +b_{\ell}\leq \bar{m}_{\ell}$. Then the ReLU activation function is equivalent to the following set of mixed-integer linear constraints~\cite{tjeng2017evaluating}:
\begin{equation}
\begin{aligned} \label{eq:MIL_NN}
& z_{\ell+1} = \max(W_\ell z_\ell + b_\ell,0) \iff \\ 
& \begin{cases}
z_{\ell+1} \geq W_\ell z_\ell + b_\ell \\ 
z_{\ell+1} \leq W_\ell z_\ell + b_\ell - \mathrm{diag}(\underline{m}_\ell) (\mathbf{1}-t_\ell) \\
z_{\ell+1} \geq 0 \\
z_{\ell+1} \leq \mathrm{diag}(\bar{m}_\ell)  t_\ell,
\end{cases} 
\end{aligned}
\end{equation}
where $t_{\ell} \in \{0,1\}^{n_{\ell+1}}$ is a vector of binary variables for the $(\ell+1)$-th activation layer. Since the input to the ReLU network is bounded in our considered problem, various methods are available to find the element-wise pre-activation bounds $\{\underline{m}_{\ell} , \bar{m}_{\ell}\}$ such as interval bound propagation~\cite{cheng2017maximum} and linear programming~\cite{wong2018provable}. By assembling the mixed-integer linear constraints~\eqref{eq:MIL_NN} for each layer, we obtain the mixed-integer linear representation of the ReLU network and also the nominal dynamics $x_+ = \hat{f}(x)$ in~\eqref{eq:uncertain_dyn}. Consequently, the composition $\hat{f}^{(i)}(x)$ for $i = 1, \cdots, k$ can be described by mixed-integer linear constraints as well.

With the given Lyapunov function candidate $V_k(x;P_{ac})$, problem~\eqref{eq:verifier_formulation} can be rewritten as
\begin{subequations}
	\label{eq:MIQP}
	\begin{align}
	\begin{split} \label{eq:MIQP_obj}
	\underset{x, y, w}{\text{maximize}} & \quad z_k^\top(y) P_{ac} z_k^\top(y) - z_k^\top(x) P_{ac} z_k^\top(x)
	\end{split} \\
	\begin{split} \label{eq:MIQP_dynanmics}
	\text{subject to} & \quad y = Ax + f_{NN}(x) + w 
	\end{split}\\
	\begin{split} \label{eq:MIQP_w_set}
	& \quad \lVert w \rVert_\infty \leq \gamma \lVert x \rVert_\infty + \delta
	\end{split}\\
	\begin{split} \label{eq:MIQP_ROI}
	& \quad x \in \mathcal{X} \setminus B_r
	\end{split}
	\end{align}
\end{subequations}
where the MIL representation of constraint~\eqref{eq:MIQP_dynanmics} and basis $z_k(x), z_k(y)$ follows from that of the ReLU network, and the MIL representation of constraints~\eqref{eq:MIQP_w_set} and~\eqref{eq:MIQP_ROI} follow from that of the $\ell_\infty$ norm. Since the objective function~\eqref{eq:MIQP_obj} is quadratic and indefinite, problem~\eqref{eq:MIQP} is a nonconvex MIQP which can be solved by Gurobi to global optimality.

\subsection{Termination guarantee}
Algorithm~\ref{alg:ACCPM} is guaranteed to find a Lyapunov function in the target set under the following assumption:
\begin{assume} \label{assump:nonempty}
	The target set $\mathcal{F}$ defined in~\eqref{eq:target_set} is full-dimensional, i.e., there exists $P_c\in \mathbb{S}^{(k+1)n_x}$ and $\epsilon > 0$ such that $\{P \in \mathbb{S}^{(k+1)n_x} \vert \lVert P - P_c \rVert_F \leq \epsilon \} \subset \mathcal{F}$ where $\lVert \cdot \rVert_F$ is the Frobenius norm. 
\end{assume}

\begin{theorem}
	\label{thm:termination}
	Under Assumption~\ref{assump:nonempty}, Algorithm~\ref{alg:ACCPM} finds a Lyapunov function in the target set $\mathcal{F}$ in at most $O(((k+1)n_x)^3/\epsilon^2)$ iterations. 
\end{theorem}

Theorem~\ref{thm:termination} states that when the target set is full-dimensional, Algorithm~\ref{alg:ACCPM} is guaranteed to find a Lyapunov function in a finite number of steps. Such a guarantee follows from the termination guarantee of the analytic center cutting-plane method considered in~\cite{sun2002analytic} and this is the main reason we require the learner to propose the analytic center of the localization sets as Lyapunov function candidates. From an optimization point of view, the sampling-based method described in Algorithm~\ref{alg:ACCPM} is equivalent to finding a feasible solution in the target set through the analytic center cutting-plane method with the verifier serving as the cutting-plane oracle. The proof of Theorem~\ref{thm:termination} follows from that of \cite[Theorem 2]{chen2021learning} since our extension of the Lyapunov function synthesis method in~\cite{chen2021learning} to the uncertain dynamical system~\eqref{eq:uncertain_dyn} does not change the convexity  of the target set. 

\begin{remark}
	When the target set is empty, we do not have a finite-step termination guarantee for the proposed algorithm. Therefore, Algorithm~\ref{alg:ACCPM} is semi-complete and in practice, we can add additional stopping criteria, e.g., maximum number of iterations, to Algorithm~\ref{alg:ACCPM}.
\end{remark}

To summarize, the sampling-based Lyapunov function synthesis method in~\cite{chen2021learning} looks for Lyapunov functions simply based on the mixed-integer representation of hybrid systems. This makes it suitable to verify the stability of neural network dynamical systems since modeling the behavior of ReLU network over a compact set through MIL constraints is straightforward. In this paper we extend this method to handle uncertain dynamical systems and apply it in a novel setting to estimate the ROA of nonlinear dynamical systems through their neural network approximations. The non-conservativeness and the finite-step termination guarantee provided by Theorem~\ref{thm:Lyap} makes this method favorable to analyze uncertain dynamical systems. However, this is obtained at the cost of the worst-case exponential complexity of the MIQP~\eqref{eq:MIQP}.

%% file: err_bound_assumption.tex
% applicability discussion
\label{sec:applicability}
In this section, we investigate the applicability of Assumption~\ref{assumption:nn approximation} and show how to estimate the error bound parameters $(\gamma, \delta)$.

\subsection{Neural network training}
The decomposition of the nonlinear map $f(x) = Ax + f_{NN}(x) + w(x)$ in~\eqref{eq:nn_approx_dyn} is motivated by two facts: first, stability verification tools~\cite{chen2021learning, yin2020stability} for neural network dynamical systems have been developed; second, powerful and easy-to-use machine learning tools are available to train $f_{NN}(x)$ to high accuracy and keep the approximation error $w(x)$ small. In training $f_{NN}(x)$, we want to obtain a neural network model that not only achieves a small approximation error but also is easy to verify. In light of the latter objective, it is desirable to increase the number of \emph{stable} ReLUs in $f_{NN}(x)$, i.e., ReLUs that always output zero or one. For these ReLUs, we do not need to introduce binary variables in the MIQP~\eqref{eq:MIQP}. As a result, any training method that can increase the number of stable ReLUs can significantly improve the complexity of the verification. As shown in~\cite{xiao2018training}, applying $\ell_1$-regularization to the training objective generally works well in reducing the complexity of verification. Other training methods with the same goal are presented in~\cite{xiao2018training} as well.

\subsection{Identification of error bound}
For a given neural network $f_{NN}(x)$ which approximates the nonlinear map $f(x)$, we estimate an error bound in the form of~\eqref{eq:approx_err_bd} based on dense sampling of the ROI $\mathcal{X}$. This method is not meant to compete with the modeling error quantification methods in the literature. Instead, we use it to provide a reasonable error bound to set up the simulation.

By assumption $f(x)$ is locally Lipschitz continuous. Let $L_f$ ($L_N$) be an upper bound on the Lipschitz constant of $f(x)$ ($Ax + f_{NN}(x)$) over the ROI $\mathcal{X}$. Since $w(x)= f(x) - Ax - f_{NN}(x)$, we can upper bound the Lipschitz constant of $w(x)$ as $L_w = L_f + L_N$. Then we sample densely from the ROI $\mathcal{X}$ and in particular the samples $\mathcal{S}_\mathcal{X} = \{x^S_i\}_{i=0}^N \subset \mathcal{X}$ form a $\epsilon$-net of $\mathcal{X}$, i.e., for all $x \in \mathcal{X}$, there exists a sample $y \in \mathcal{S}_\mathcal{X}$ such that $\lVert x - y \rVert \leq \epsilon$. We define the admissible set of $(\gamma, \delta)$ as 
\begin{equation*}
\Gamma =\{(\gamma, \delta) \vert \lVert w(x) \rVert_\infty  \leq \gamma \lVert x \rVert_\infty + \delta, \forall x \in \mathcal{S}_\mathcal{X} \}
\end{equation*} 
and any pair $(\gamma^\prime, \delta^\prime) \in \Gamma$ gives an estimate of the upper bound in~\eqref{eq:approx_err_bd}. Since we want to keep $(\gamma^\prime, \delta^\prime)$ small, we let $\delta^\prime = \min \delta \text{ s.t. } (\gamma^\prime, \delta) \in \Gamma$. Note that there are infinitely many pairs in the admissible set and decreasing one parameter would increase the other one. 

%How to trade-off between the magnitudes of $\gamma^\prime$ and $\delta^\prime$ is left for discussion in Section~\ref{}.

Given any admissible pair $(\gamma^\prime, \delta^\prime)$, for all $x \in \mathcal{X}$ we have 
\begin{equation} \label{eq:err_bd_approx}
\begin{aligned}
\lVert w(x) \rVert_\infty &\leq \lVert w(x) - w(y)\rVert_\infty + \lVert w(y) \rVert_\infty  \\
&\leq L_w \lVert x - y \rVert_\infty + \gamma^\prime \lVert y \rVert_\infty  + \delta^\prime\\
& \leq L_w \lVert x - y \rVert_\infty + \gamma^\prime \lVert y - x \rVert_\infty + \gamma^\prime \lVert x \rVert_\infty  + \delta^\prime \\
& \leq \gamma^\prime \lVert x \rVert_\infty  + \delta^\prime + (L_w + \gamma^\prime) \epsilon
\end{aligned}
\end{equation}
where $y \in \mathcal{S}_\mathcal{X}$ is chosen such that $\lVert x - y \rVert \leq \epsilon$. Therefore, $\gamma = \gamma^\prime, \delta = \delta^\prime + (L_w + \gamma^\prime) \epsilon$ gives a valid upper bound for $\lVert w(x) \rVert_\infty$ over the ROI $\mathcal{X}$ if we assume $L_f$ is known. Otherwise, inequality~\eqref{eq:err_bd_approx} indicates that by choosing $\epsilon$ small enough, the admissible pair $(\gamma^\prime, \delta^\prime)$ obtained from sampling gives a good estimate of a deterministic error bound. 

Since every pair from the admissible set $\Gamma$ generates an error bound estimate, we can combine $K$ admissible $(\gamma^\prime, \delta^\prime)$ pairs to obtain a tighter error bound as
\begin{equation} \label{eq:PWA_err_bd}
	\lVert w \rVert_\infty \leq \gamma_i^\prime \lVert x \rVert_\infty + \delta_i^\prime, \quad i = 1,\cdots, K
\end{equation}
from the sampling. One special, interesting choice is setting $\gamma^\prime = 0$ since in this case constraint~\eqref{eq:MIQP_w_set} in the MIQP becomes $\lVert w \rVert_{\infty} \leq \delta^\prime$ and is convex.

\begin{figure}[htb!]
	\centering
	\includegraphics[width = 0.6\columnwidth]{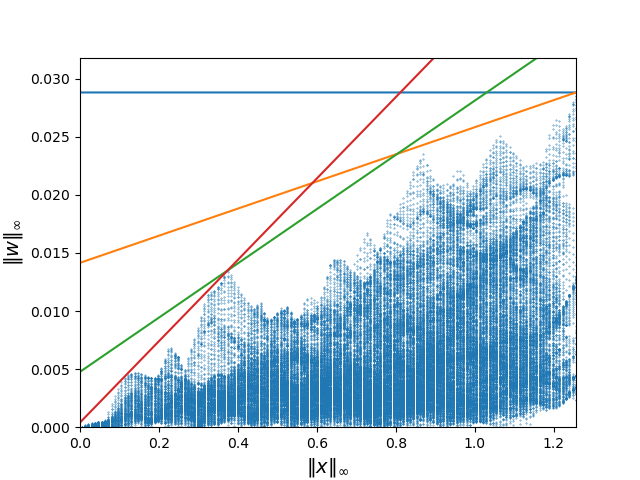}
	\caption{The neural network approximation errors are marked by the blue dots for a dense sample set over the ROI. The affine error bounds are denoted by the blue, orange, green, and red  lines corresponding to $(\gamma, \delta) = $ $(0, 0.0288)$, $(1.167, 0.0141)$, $(0.0233, 4.767 \times 10^{-3})$, $(0.0350, 4.026 \times 10^{-4})$, respectively.}
	\label{fig:err_bounds}
\end{figure}

%% file: simulation.tex
% numerical examples
\label{sec:simulation}
We consider a $2$- and $3$-dimensional nonlinear system from the literature and train ReLU networks to approximate their dynamics. The approximation errors are estimated from sampling to generate the uncertain neural network dynamical systems, for which our proposed algorithm is called to synthesize robust Lyapunov functions. All experiments are run on an Intel Core i7-6700K CPU.

\subsection{2-D rational system}
Consider the $2$ dimensional discrete-time rational system adapted from~\cite[Example 2]{coutinho2013local}:
\begin{equation} \label{eq:rational}
x_1^+ = x_1 - \frac{x_1 + x_2^3}{1+ x_1^2}, \quad x_2^+ = x_2 + \frac{x_1^3 - 0.25x_2}{1 + x_2^2}.
\end{equation}
and denote the dynamics in~\eqref{eq:rational} as $x_+ = f(x)$. For the uncertain dynamics decomposition~\eqref{eq:uncertain_dyn}, we choose the Jacobian of $f(x)$ at the origin as matrix $A = [0 \ 0; 0 \ 0.75]$. Then we sample $22500$ states uniformly from the box $\mathcal{R} = \{x \in \mathbb{R}^2 \vert \lVert x \rVert_\infty \leq 1.5 \}$ and train a fully-connected ReLU network with the structure $2-100-100-100-2$ (three hidden layers with $100$ neurons each) with Keras~\cite{chollet2015keras} and the Adam optimizer~\cite{kingma2014adam} in $100$ epochs to approximate the error dynamics $f(x) - Ax$. Since the approximation error bound~\eqref{eq:approx_err_bd} is written in terms of the $\ell_\infty$ norm, the training loss function is set to the mean absolute error. Furthermore, in order to alleviate the practical computational cost of the MIQP~\eqref{eq:MIQP} in the analysis phase, we add $\ell_1$ regularization with weight  $10^{-4}$ during training. 

\begin{figure}[htb!]
\centering
\includegraphics[width = 0.55 \columnwidth]{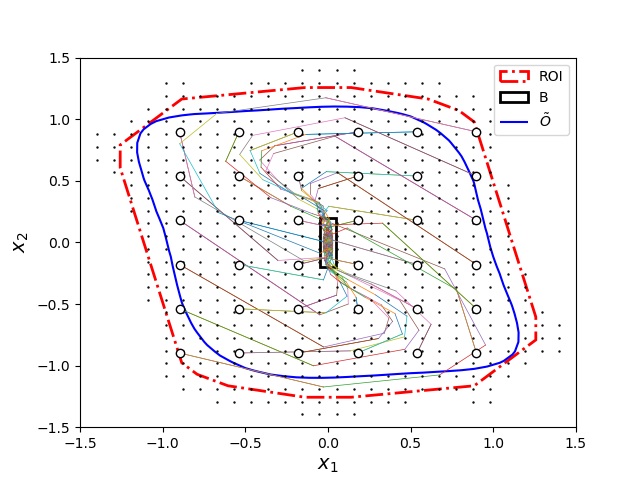}
\caption{Estimate ROA of system~\eqref{eq:rational} through neural network approximation and synthesizing robust Lyapunov function. The estimate ROA $\tilde{O}$ is marked by the blue curve and the ROI is given by the red polytope with dashed lines. The set $B$ is a rectangle around the origin which is excluded from the search space of the verifier.}
\label{fig:rational_ROA}
\end{figure}

We construct the ROI $\mathcal{X}$ by uniformly sampling initial states from the region $\mathcal{R}$ and simulate the trajectories of the system~\eqref{eq:rational} for $50$ steps. Then, the convex hull of the initial condition samples from which the trajectories converge to the origin is chosen as the reference ROI $\mathcal{X}_\text{ref} := \{ x \vert A_\text{ref} x \leq b_\text{ref} \}$. We tune the ROI by $\mathcal{X} = \tau \mathcal{X}_\text{ref} := \{x\vert A_\text{ref} \leq \tau b_\text{ref}\}$ with $0 < \tau \leq 1.0$. In this example, we set the scaling parameter $\tau = 0.9$. The sampled convergent initial states are marked as the black dots in Fig.~\ref{fig:rational_ROA} where the ROI is also plotted. 

After training the neural network and fixing the ROI, we estimate the neural network approximation error by sampling a uniformly spaced grid from the ROI. The grid consists of $72306$ state samples which form an $\epsilon$-net of the ROI with $\epsilon \approx 0.005$. We plot the $(\lVert x \rVert_\infty, \lVert w \rVert_\infty)$ pairs for all the samples in Fig.~\ref{fig:err_bounds} and our estimated error bound with $(\gamma, \delta)$ gives an affine upper bound over all the sampled graph of $(\lVert x \rVert_\infty, \lVert w \rVert_\infty)$. Selected values of $(\gamma, \delta)$ and their corresponding affine bounds are plotted in Fig.~\ref{fig:err_bounds}. Although each pair $(\gamma, \delta)$ gives an estimate on the bound of the neural network approximation error, formulating a concave piecewise affine upper bound by combining different values of $(\gamma, \delta)$ is tighter than using a single one as discussed in Section~\ref{sec:applicability}. As shown in~\eqref{eq:err_bd_approx}, if the sampling is dense enough, the estimated error bounds from samples is close to the true bounds. In this section, we apply estimated error bounds from sampling for simplicity. 

Finally, we run Algorithm~\ref{alg:ACCPM} to find a robust Lyapunov function for the considered neural network approximation system with the error bound given by the concave bound shown in Fig.~\ref{fig:err_bounds} using four pairs of $(\gamma, \delta)$ values. Let $B$ denote the neighborhood that is excluded from the search space in place of the $\ell_\infty$ norm ball $B_r$ in~\eqref{eq:robust_Lyap_cond}. For this example, we choose $B = \{x \in \mathbb{R}^2 \vert -0.05 \leq x_1 \leq 0.05, -0.2 \leq x_2 \leq 0.2\}$ as a rectangle according to the simulated uncertain dynamical system~\eqref{eq:uncertain_dyn}. Then we set the order of the Lyapunov function candidates to be $k = 1$ in~\eqref{eq:V_order} and initialize Algorithm~\ref{alg:ACCPM} with an empty sample set. Algorithm~\ref{alg:ACCPM} terminates in $6$ iterations with a total solver time of $748$ seconds. It successfully finds a robust Lyapunov function $V_1(x;P)$ with $P \in \mathbb{S}^{4}_{++}$. The resulting estimate of the ROA $\tilde{\mathcal{O}}$ which is the largest sublevel set of $V_1(x;P)$ is plotted in Fig.~\ref{fig:rational_ROA} together with several simulated trajectories of the uncertain system~\eqref{eq:uncertain_dyn}. We observe that the estimate ROA provided by $V_1(x;P)$ is close to the true ROA estimated by the samples. 

As a comparison, we run Algorithm~\ref{alg:ACCPM} with error bound $(\gamma , \delta) = (0, 0.0288)$ which corresponds to the blue line in Fig~\ref{fig:err_bounds}. Since $\gamma = 0$, the constraint~\eqref{eq:MIQP_w_set} becomes convex at the cost of looser characterization of the approximation errors. With the same setup, Algorithm~\ref{alg:ACCPM} finds a robust Lyapunov function in $17$ iterations with the total solver time $3162$ seconds. The synthesized robust Lyapunov function generates an estimate ROA similar to that shown in Fig~\ref{fig:rational_ROA} with about $4$ times the running time.

\subsection{3-D polynomial system}
Consider the $3$-dimensional continuous-time polynomial system in~\cite[Example B]{doban2017computation}:
\begin{equation} \label{eq:3d_system}
\begin{bmatrix}
\dot{x}_1 \\ \dot{x}_2 \\ \dot{x}_3
\end{bmatrix} =  \begin{bmatrix}
x_1(x_1^2 + x_2^2 - 1) 	- x_2(x_3^2 + 1) \\
x_2(x_1^2 + x_2^2 - 1) 	+ x_1(x_3^2 + 1)  \\
10 x_3(x_3^2 - 1)
\end{bmatrix}.
\end{equation}
We apply the Euler discretization method with sampling time $dt = 0.1$ seconds to discretize the dynamics and choose the Jacobian of the discretized dynamics at the origin as the $A$ matrix. Then we apply a $3-100-100-100-3$ fully-connected ReLU network to approximate the discrete-time dynamics over the box $\mathcal{R} = \{x \in  \mathbb{R}^3 \vert \lVert x \rVert_\infty \leq 1 \}$. The neural network is trained with $125000$ samples uniformly spaced in $\mathcal{R}$ by the same training method described in the $2$-dimensional example. The ROI is obtained similarly through simulation and we estimate the neural network approximation error bound from sampling. The samples form an $\epsilon$-net in the ROI with $\epsilon \approx 0.01$ and the error bound $(\gamma, \delta) = (0, 0.0165)$ is estimated from sampling. 

\begin{figure}
	\centering
	\begin{subfigure}[b]{0.48\columnwidth}            
		\includegraphics[width= \textwidth]{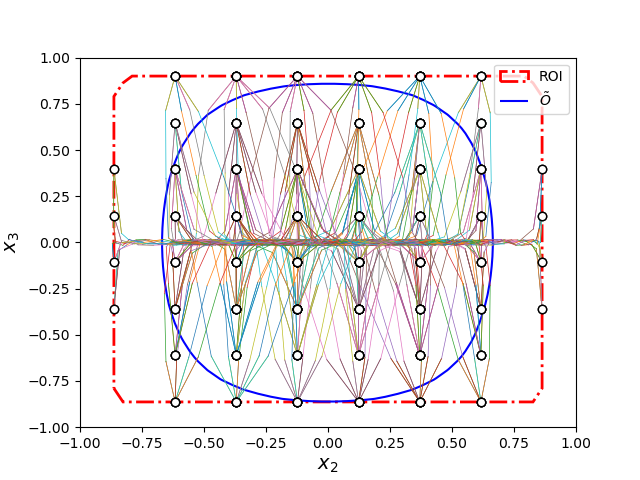}
		\caption{Slice at $x_1 = 0$}
		\label{fig:slice_x_1}
	\end{subfigure}
 	\hfil
	\begin{subfigure}[b]{0.48\columnwidth}
		\centering
		\includegraphics[width= \textwidth]{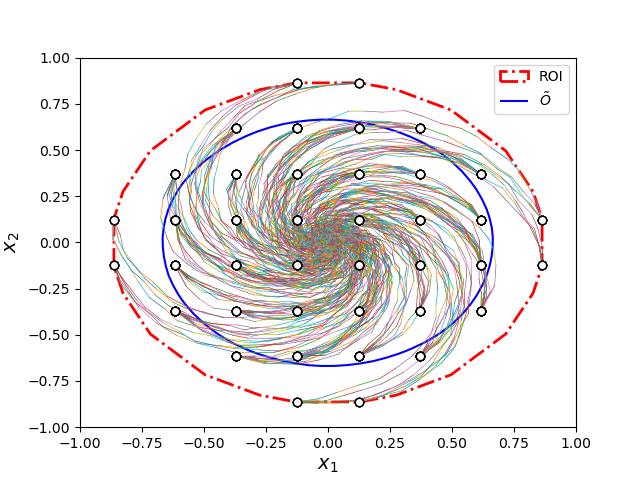}
		\caption{Slice at $x_3 = 0$}
		\label{fig:slice_x_3}
	\end{subfigure}
	\caption{Estimated ROA $\tilde{\mathcal{O}}$ (blue curve) given by the robust Lyapunov function and the ROI (red polytope with dashed lines) on slices $x_1 = 0$ and $x_3 = 0$. Simulated trajectories of the uncertain neural network dynamical system and their initial states (white dots) are projected to the corresponding slices for plotting. }
	\label{fig:ROA_slices}
\end{figure}

We run Algorithm~\ref{alg:ACCPM} for the uncertain neural network dynamical system with the exclusion of $B = \{x \in \mathbb{R}^3 \vert \lVert x \rVert_\infty \leq 0.05\}$ from the ROI. The sample set is initialized as an empty set. With order $k = 2$, the algorithm terminates in one iteration with a total solver time of $10$ seconds. Upon termination, a robust Lyapunov function $V_2(x;P)$ is found whose largest sublevel set is shown in Fig~\ref{fig:ROA_slices} together with the ROI at slices $x_1 = 0$ and $x_3 = 0$. Simulated trajectories of the uncertain neural network dynamical system starting from uniformly spaced initial conditions inside the ROI are plotted in Fig.~\ref{fig:ROA_slices} to illustrate the system dynamics. 

%The estimate ROA given by $V_2(x;P)$ is similar in shape to the estimated ROA found in~\cite[Example B]{doban2017computation} for the continuous-time system~\eqref{eq:3d_system}.

%% file: conclusion.tex
% conclusion
\label{sec:conclusion}
Finding Lyapunov functions for nonlinear dynamical systems and estimating their region of attraction relies on deep insights from experts and requires a case-by-case analysis. In this paper, we move a step towards automating Lyapunov function synthesis for nonlinear systems. Our method starts with approximating the map of a dynamical system by a piecewise linear neural network. Assuming that the resulting approximation error can be bounded, we then develop a counterexample-guided method to synthesize a Lyapunov function that is robust to the approximation error.

While in practice the approximation error can be made arbitrarily small, computing deterministic bounds on it can be challenging, especially in high dimensions. In future work, we will explore efficient ways to bound the approximation error.

%% file: main.bbl
% Generated by IEEEtran.bst, version: 1.14 (2015/08/26)
\begin{thebibliography}{10}
\providecommand{\url}[1]{#1}
\csname url@samestyle\endcsname
\providecommand{\newblock}{\relax}
\providecommand{\bibinfo}[2]{#2}
\providecommand{\BIBentrySTDinterwordspacing}{\spaceskip=0pt\relax}
\providecommand{\BIBentryALTinterwordstretchfactor}{4}
\providecommand{\BIBentryALTinterwordspacing}{\spaceskip=\fontdimen2\font plus
\BIBentryALTinterwordstretchfactor\fontdimen3\font minus
  \fontdimen4\font\relax}
\providecommand{\BIBforeignlanguage}[2]{{%
\expandafter\ifx\csname l@#1\endcsname\relax
\typeout{** WARNING: IEEEtran.bst: No hyphenation pattern has been}%
\typeout{** loaded for the language `#1'. Using the pattern for}%
\typeout{** the default language instead.}%
\else
\language=\csname l@#1\endcsname
\fi
#2}}
\providecommand{\BIBdecl}{\relax}
\BIBdecl

\bibitem{haddad2011nonlinear}
W.~M. Haddad and V.~Chellaboina, \emph{Nonlinear dynamical systems and control:
  a {L}yapunov-based approach}.\hskip 1em plus 0.5em minus 0.4em\relax
  Princeton university press, 2011.

\bibitem{giesl2015review}
P.~Giesl and S.~Hafstein, ``Review on computational methods for {L}yapunov
  functions,'' \emph{Discrete \& Continuous Dynamical Systems-B}, vol.~20,
  no.~8, p. 2291, 2015.

\bibitem{berkenkamp2016safe}
F.~Berkenkamp, R.~Moriconi, A.~P. Schoellig, and A.~Krause, ``Safe learning of
  regions of attraction for uncertain, nonlinear systems with gaussian
  processes,'' in \emph{2016 IEEE 55th Conference on Decision and Control
  (CDC)}.\hskip 1em plus 0.5em minus 0.4em\relax IEEE, 2016, pp. 4661--4666.

\bibitem{wang2018permissive}
L.~Wang, D.~Han, and M.~Egerstedt, ``Permissive barrier certificates for safe
  stabilization using sum-of-squares,'' in \emph{2018 Annual American Control
  Conference (ACC)}.\hskip 1em plus 0.5em minus 0.4em\relax IEEE, 2018, pp.
  585--590.

\bibitem{richards2018lyapunov}
S.~M. Richards, F.~Berkenkamp, and A.~Krause, ``The {L}yapunov neural network:
  Adaptive stability certification for safe learning of dynamical systems,'' in
  \emph{Conference on Robot Learning}.\hskip 1em plus 0.5em minus 0.4em\relax
  PMLR, 2018, pp. 466--476.

\bibitem{giesl2016approximation}
P.~Giesl, B.~Hamzi, M.~Rasmussen, and K.~N. Webster, ``Approximation of
  {L}yapunov functions from noisy data,'' \emph{arXiv preprint
  arXiv:1601.01568}, 2016.

\bibitem{boffi2020learning}
N.~M. Boffi, S.~Tu, N.~Matni, J.-J.~E. Slotine, and V.~Sindhwani, ``Learning
  stability certificates from data,'' \emph{arXiv preprint arXiv:2008.05952},
  2020.

\bibitem{ogunmolu2016nonlinear}
O.~Ogunmolu, X.~Gu, S.~Jiang, and N.~Gans, ``Nonlinear systems identification
  using deep dynamic neural networks,'' \emph{arXiv preprint arXiv:1610.01439},
  2016.

\bibitem{chen2021learning}
S.~Chen, M.~Fazlyab, M.~Morari, G.~J. Pappas, and V.~M. Preciado, ``Learning
  {L}yapunov functions for hybrid systems,'' in \emph{Proceedings of the 24nd
  ACM International Conference on Hybrid Systems: Computation and Control},
  2021.

\bibitem{yin2020stability}
H.~Yin, P.~Seiler, and M.~Arcak, ``Stability analysis using quadratic
  constraints for systems with neural network controllers,'' \emph{arXiv
  preprint arXiv:2006.07579}, 2020.

\bibitem{ahmed2020automated}
D.~Ahmed, A.~Peruffo, and A.~Abate, ``Automated and sound synthesis of
  {L}yapunov functions with smt solvers,'' in \emph{International Conference on
  Tools and Algorithms for the Construction and Analysis of Systems}.\hskip 1em
  plus 0.5em minus 0.4em\relax Springer, 2020, pp. 97--114.

\bibitem{abate2020formal}
A.~Abate, D.~Ahmed, M.~Giacobbe, and A.~Peruffo, ``Formal synthesis of
  {L}yapunov neural networks,'' \emph{IEEE Control Systems Letters}, vol.~5,
  no.~3, pp. 773--778, 2020.

\bibitem{chang2020neural}
Y.-C. Chang, N.~Roohi, and S.~Gao, ``Neural {L}yapunov control,'' \emph{arXiv
  preprint arXiv:2005.00611}, 2020.

\bibitem{ahmadi2008non}
A.~A. Ahmadi and P.~A. Parrilo, ``Non-monotonic {L}yapunov functions for
  stability of discrete time nonlinear and switched systems,'' in \emph{2008
  47th IEEE Conference on Decision and Control}.\hskip 1em plus 0.5em minus
  0.4em\relax IEEE, 2008, pp. 614--621.

\bibitem{bobiti2016sampling}
R.~Bobiti and M.~Lazar, ``A sampling approach to finding {L}yapunov functions
  for nonlinear discrete-time systems,'' in \emph{2016 European Control
  Conference (ECC)}.\hskip 1em plus 0.5em minus 0.4em\relax IEEE, 2016, pp.
  561--566.

\bibitem{nesterov1995cutting}
Y.~Nesterov, ``Cutting plane algorithms from analytic centers: efficiency
  estimates,'' \emph{Mathematical Programming}, vol.~69, no.~1, pp. 149--176,
  1995.

\bibitem{atkinson1995cutting}
D.~S. Atkinson and P.~M. Vaidya, ``A cutting plane algorithm for convex
  programming that uses analytic centers,'' \emph{Mathematical Programming},
  vol.~69, no. 1-3, pp. 1--43, 1995.

\bibitem{boyd2004convex}
S.~Boyd and L.~Vandenberghe, \emph{Convex optimization}.\hskip 1em plus 0.5em
  minus 0.4em\relax Cambridge university press, 2004.

\bibitem{belotti2013mixed}
P.~Belotti, C.~Kirches, S.~Leyffer, J.~Linderoth, J.~Luedtke, and A.~Mahajan,
  ``Mixed-integer nonlinear optimization,'' \emph{Acta Numerica}, vol.~22,
  p.~1, 2013.

\bibitem{belotti2009branching}
P.~Belotti, J.~Lee, L.~Liberti, F.~Margot, and A.~W{\"a}chter, ``Branching and
  bounds tightening techniques for non-convex minlp,'' \emph{Optimization
  Methods \& Software}, vol.~24, no. 4-5, pp. 597--634, 2009.

\bibitem{gurobi}
\BIBentryALTinterwordspacing
{Gurobi Optimization, LLC}, ``{Gurobi Optimizer Reference Manual},'' 2021.
  [Online]. Available: \url{https://www.gurobi.com}
\BIBentrySTDinterwordspacing

\bibitem{tjeng2017evaluating}
V.~Tjeng, K.~Xiao, and R.~Tedrake, ``Evaluating robustness of neural networks
  with mixed integer programming,'' \emph{arXiv preprint arXiv:1711.07356},
  2017.

\bibitem{cheng2017maximum}
C.-H. Cheng, G.~N{\"u}hrenberg, and H.~Ruess, ``Maximum resilience of
  artificial neural networks,'' in \emph{International Symposium on Automated
  Technology for Verification and Analysis}.\hskip 1em plus 0.5em minus
  0.4em\relax Springer, 2017, pp. 251--268.

\bibitem{wong2018provable}
E.~Wong and Z.~Kolter, ``Provable defenses against adversarial examples via the
  convex outer adversarial polytope,'' in \emph{International Conference on
  Machine Learning}, 2018, pp. 5286--5295.

\bibitem{sun2002analytic}
J.~Sun, K.-C. Toh, and G.~Zhao, ``An analytic center cutting plane method for
  semidefinite feasibility problems,'' \emph{Mathematics of Operations
  Research}, vol.~27, no.~2, pp. 332--346, 2002.

\bibitem{xiao2018training}
K.~Y. Xiao, V.~Tjeng, N.~M. Shafiullah, and A.~Madry, ``Training for faster
  adversarial robustness verification via inducing relu stability,''
  \emph{arXiv preprint arXiv:1809.03008}, 2018.

\bibitem{coutinho2013local}
D.~Coutinho and C.~E. de~Souza, ``Local stability analysis and domain of
  attraction estimation for a class of uncertain nonlinear discrete-time
  systems,'' \emph{International Journal of Robust and Nonlinear Control},
  vol.~23, no.~13, pp. 1456--1471, 2013.

\bibitem{chollet2015keras}
\BIBentryALTinterwordspacing
F.~Chollet \emph{et~al.} (2015) Keras. [Online]. Available:
  \url{https://github.com/fchollet/keras}
\BIBentrySTDinterwordspacing

\bibitem{kingma2014adam}
D.~P. Kingma and J.~Ba, ``Adam: A method for stochastic optimization,''
  \emph{arXiv preprint arXiv:1412.6980}, 2014.

\bibitem{doban2017computation}
A.~I. Doban and M.~Lazar, ``Computation of {L}yapunov functions for nonlinear
  differential equations via a massera-type construction,'' \emph{IEEE
  Transactions on Automatic Control}, vol.~63, no.~5, pp. 1259--1272, 2017.

\end{thebibliography}
